# Minima hopping: Searching for the global minimum of the potential energy surface of complex molecular systems without invoking thermodynamics

Stefan Goedecker

Institut für Physik
Universität Basel, Switzerland

## 1 Abstract

A method is presented that can find the global minimum of very complex condensed matter systems. It is based on the simple principle of exploring the configurational space as fast as possible and of avoiding revisiting known parts of this space. Even though it is not a genetic algorithm, it is not based on thermodynamics. The efficiency of the method depends strongly on the type of moves that are used to hop into new local minima. Moves that find low-barrier escape-paths out of the present minimum lead generally into low energy minima.

## 2 Introduction

Finding the global minimum of the potential energy surface of a complex system is a central problem in physics, chemistry and biology. For a periodic system the global minimum gives the crystalline ground state structure of a solid, for a non-periodic systems it determines the geometric ground state configuration of molecules. Many complex biomolecules exist in nature whose structure is unknown. In the case of proteins the ground state is called the native state and the theoretical determination of this native state is considered to be one of the major challenges of modern biology.

Because the experimental determination of the geometric ground state of complex systems is very difficult, simulation is a promising candidate for the theoretical determination of the structure of such systems. Several unknown alloy structures were for instance recently found by simulation methods [1]. Finding the ground state structure of large systems is computationally very expensive and has only become possible on the latest generation of computers. The fundamental reason why finding the global minimum is so expensive is that the number of local minima increases exponentially with respect to the number of atoms in the system. Initially the system is trapped in one of these local minima, and in order to reach the global minimum the system has to travel through many intermediate local minima basins and it has to overcome the barriers separating the local minima. A basin is, by the conventional definition, a certain part of the configurational space around a minimum of the potential energy surface. More precisely, a basin contains all the configurations that will relax into this minimum using simple small-step downhill relaxations. We will denote as a super-basin the union of several neighboring basins. If one can arrive from any point in such a super-basin into the lowest minimum of this super-basin without crossing barriers that are very high compared to the average difference in energy between local minima, it will be called a funnel [2]. The large majority of algorithms for finding the global minimum in physical systems is based on thermodynamic principles. Because of the ubiquitous presence of the Boltzmann factor $\exp(-\Delta E/k_b T)$ in these algorithms climbing and crossing over high barriers that separate for instance different funnels is a rare event that frequently does not happen during the available computer time.

# 3 Finding the global minimum without thermodynamics

With the exception of genetic algorithms [3], most standard algorithms such as simulated annealing [4], basin hopping [4] and multi-canonical methods [5] are based on thermodynamic principles. A Markov process based on the Metropolis algorithm with a Boltzmann factor leads finally to a thermodynamic distribution. At sufficiently low temperature the ground state configuration will be the dominant configuration and hence the problem is solved in principle. Unfortunately thermodynamics does not tell us anything about how fast the thermodynamic equilibrium distribution is obtained and as a matter of fact it can be extremely slow. Consequently, global minimization strategies based on thermodynamics are of questionable value.

Instead of invoking thermodynamic principles, the basic principle in global optimization should be that one strives to explore as fast as possible the low energy part of the configurational space. Revisiting configurations is obviously detrimental for finding new configurations. All ordinary Monte Carlo like algorithms have the tendency to revisit many times neighboring configurations that are close in energy, if other configurations can only be reached by crossing high barriers. As a matter of fact the simulation can just jump back and forth between two such configurations for a very long time.

The problem of repeated visits of certain configurations has already been recognized by many researches. One remedy that was proposed is flooding [6]. The principle of flooding is simple. In basins that were already visited during the simulation, the potential is lifted and consequently it is less likely that a configuration in the same region of the continuous configuration space will be accepted in a future Monte Carlo step. The problem with this approach is that it is very difficult to determine exactly where the potential should be lifted. Ideally it should be done over the whole basin, but the determination of the basin boundaries would be prohibitively expensive computationally. Because of this problem flooding has not found widespread use.

Because it is so difficult to determine the shape of a basin that has to be flooded, another version of flooding has been proposed recently. The flooding is not done in the very high dimensional configurational space, but in a low dimensional space spanned by some suitably defined order parameters of the system [7]. The difficulties of determining the shape of the basin to be flooded is not really eliminated in this method, but it is alleviated since the space dimension is much lower.

There is still a second more profound problem with flooding. For a simple system, where essentially all basins can be visited during the simulation, energy surface flooding is a method bound for success. Successively all the basins will be flooded until the system finds its way into the global minimum. In a more realistic high dimensional setting with a very large number of basins, things can be different. Imagine a transition basin that is a low energy transition region among several funnels. We postulate that direct transitions between these funnels are unlikely since they are separated by high barriers from the other funnels, with the exception of the transition basin. In order to find the global minimum the simulation has to jump into all the different funnels and each jump requires a visit of the transition basin. Flooding the transition basin will make such jumps more unlikely and will thus slow down the search for the global minimum.

What is needed is a strategy that limits repeated visits, but does not penalize crossings through important transition basins. As will be shown, this can be achieved by doing more violent escape moves out of the current basin if this basin has already been visited. This then gives rise to the minima hopping method.

## 4 Minima hopping

The Minima Hopping Method (MHM) consists of an inner part that performs jumps into the local minimum of another basin and an outer part that will accept or reject this new local minimum. The acceptance/rejection is done by simple thresholding, , i.e. the step is accepted if the energy of the new local minimum $E_{new}$ rises by less than $E_{diff}$ compared to the current energy $E_{cur}$. The parameter $E_{diff}$ is continuously adjusted during the simulation in such a way, that half of the moves are accepted and half are rejected. This outer part introduces a preference for steps that go down in energy. However if the inner part proposes only steps that go up in energy, such steps will finally also be accepted after $E_{diff}$ has been sufficiently increased after many rejections.

The inner part consists of an escape step away from the current local minimum followed by a geometry relaxation into the closest local minimum. The geometry relaxation is done by a combination of standard steepest descent and conjugate gradient methods. The escape step is done by a short molecular dynamics simulation that starts from the current minimum. The atoms have a Boltzmann velocity distribution such that their kinetic energy is equal to $E_{kinetic}$. The system has thus sufficient energy to cross over a barrier of height less than $E_{kinetic}$ measured relative to the current minimum. If $E_{kinetic}$ is small, one will usually fall back into the current minimum, if it is sufficiently big, one will most likely be ejected from the current basin and end up in a different minimum. The molecular dynamics simulation is stopped as soon as the potential energy has crossed *mdmin* maxima and reached the *mdmin*-th minimum along the trajectory. At this point the geometry relaxation starts. Different molecular dynamics simulations (with different random velocity Boltzmann distributions) are started until the geometry relaxation will give a new local minimum.

Three different cases can be distinguished for the outcome of this inner part. In the first case the geometry relaxation will give back the current local minimum that was used as the starting point for the displacements. The second case is that the new minimum is one that has already previously been visited during the simulation (i.e. that has been accepted in a previous outer acceptance/rejection step). The third case is that the minimum is a new one that has not been visited previously. The third case is the desirable one, since it will result in the exploration of new configurations.

Since it is desirable to explore new configurations one might think that it is most advantageous to choose $E_{kinetic}$ very large. This would be true if the height of the new local minimum "behind" the barrier was independent of the barrier height. If the height of the barrier is correlated with the height of the new local minimum, the situation is entirely different. Since the density of states of the local minima (i.e the number of local minima per unit energy) grows exponentially with energy, we would have to explore a much larger number of local minima if we allow high barriers to be crossed. As described in detail later, we found a very clear correlation between the barrier height and the height of the new minimum. For this reason one has to make a compromise in the choice of $E_{kinetic}$ to obtain the shortest possible overall simulation time. If $E_{kinetic}$ is too large, the discrete space of the local minima that has to be explored becomes too large; if it is too small, one has to start a very large number of molecular dynamics simulations until an escape path is possibly found. In practice, it has turned out that adjusting $E_{kinetic}$ dynamically during the simulation in such a way that half of all molecular dynamics simulations will bring us into new basins is close to optimal. It has to be stressed that molecular dynamics is very powerful to find low barrier paths into other basins. Out of several alternatives we examined, it gave by far the lowest barriers. If another method was used, it might be necessary to allow for a much large fraction of the escape trials to fall back into the current minimum in order to avoid large values of $E_{kinetic}$.

The value of $E_{kinetic}$ is not only increased if one falls back into the current minimum, but also if we get into a another minimum that was already visited. If the simulation starts to "walk" around between minima that were already visited, $E_{kinetic}$ will be increased at each molecular dynamics restart until enough energy is available to cross into a new unexplored region of the discrete configurational space. This procedure does not restrict repeated visits to strategic transition basins between several basins. If such a minima of a transition basin is revisited after extensively and repeatedly exploring some funnel, the simulation will arrive at the transition state with a large value of $E_{kinetic}$ and it will therefore go on to other states that are far away from this transition basin. This is a very desirable effect.

In order to decide whether $E_{kinetic}$ will be increased or decreased we have to keep track of all the minima visited previously. Introducing a history evidently destroys the Markovian character of the simulation. This however is no disadvantage since we do not intend to generate any thermodynamic distribution.

A flowchart of the algorithm is shown below. It contains 5 parameters. $\alpha_1$ and $\alpha_2$ determine how rapidly $E_{diff}$ is increased or decreased in the case where a new configuration is rejected or accepted. $\beta_1$, $\beta_2$ and $\beta_3$ determine how rapidly $E_{kinetic}$ is modified depending on the outcome of an escape trial.

```
        initialize a current minimum 'Mcurrent'

MDstart
      ESCAPE  TRIAL PART
        start a MD trajectory with kinetic energy Ekinetic
        from current minimum 'Mcurrent'. Once potential energy
        reaches the mdmin-th minimum along the trajectory stop MD and
        optimize geometry to find the closest local minimum 'M'

        if ('M' equals 'Mcurrent') then
           Ekinetic = Ekinetic*beta1  (beta1 > 1)
           goto MDstart
        else if ('M' equals  a minimum visited previously) then
           Ekinetic = Ekinetic*beta2  (beta2 > 1)
           goto MDstart
        else if ('M' equals  a new minimum ) then
           Ekinetic = Ekinetic*beta3  (beta3 < 1)
        endif

      DOWNWARD PREFERENCE PART
        if ( energy('M') - energy('Mcurrent') < Ediff ) then
           accept new minimum: 'Mcurrent' = 'M'
           add 'Mcurrent' to history list
           Ediff = Ediff*alpha1 (alpha1 < 1)
        else if rejected
           Ediff = Ediff*alpha2 (alpha2 > 1)
        endif
```

```
            goto MDstart
```

In the simulation that will be presented later we used the values $\beta_1 = \beta_2 = 1/\beta_3 = 1.02$ and $\alpha_1 = 1/\alpha_2 = 1.05$. Increasing $\beta_2$ or $\beta_3$ leads to a less thorough search. With the above values the simulations never failed to find the minimum. With increased values the global minimum is found faster in the cases where it is found, but there are also cases where it can be missed. After the system has explored the low energy configurations it starts to explore higher energy regions. This is a consequence of the fact that $E_{kinetic}$ is increased whenever a known low energy configuration is revisited. Finally the system will explode. At this point the simulation should be stopped because the global minimum was most likely found if the search was thorough enough. When the system switches over from going on the average down in energy to going up depends again on the values of $\beta_2$ and $\beta_3$. For a coarse search the system will soon start going up in energy after it has visited a small fraction of the total local minima, for a fine search it will start going up after a very long time and after having visited a large fraction of all minima.

# 5 Significance of the Bell-Evans-Polanyi principle for global optimization

The Bell-Evans-Polanyi principle [8] states that highly exothermic chemical reactions have a low activation energy. In our language this means that it is more likely to find a low energy local minimum if one crosses from the current basin over a low barrier into a new basin than if one overcomes a high barrier. As a matter of fact, the Bell-Evans-Polanyi principle is more likely to be well satisfied in the context of global optimization than in the context of single chemical reactions. In the context of global optimization it has to hold only in an average sense and not for each individual barrier crossing. A difference is that the Bell-Evans-Polanyi principle is a statement about the true physical transition state, i.e. about the lowest barrier separating two basins. In the context of global optimization one crosses usually not from one basin to another via the exact transition state but over a higher energy barrier. Nevertheless the Bell-Evans-Polanyi principle should hold equally well for reaction paths that are not going through the exact physical transition state. The basis assumption of the Bell-Evans-Polanyi principle, namely that the potential energy landscape along the reaction path can be approximated by two parabolas centered at the two minima (Figure 1) is valid for any reasonable reaction path that does not involve enormous barriers.

Numerical results clearly show that the Bell-Evans-Polanyi principle holds for the reaction paths of ordinary escape trials. Table 1 shows results for a system that has been studied extensively, the 38 atom Lennard Jones cluster [9]. These studies have shown that the system has two funnels. The funnel containing the ground state contains only a relatively small number of other minima. The second funnel on the other hand contains a lot of minima. Therefore practically all runs first fall into this broad funnel. For this reason we have actually chosen the deepest minimum of this funnel, which is the second deepest overall minimum, as the starting point of our simulation.

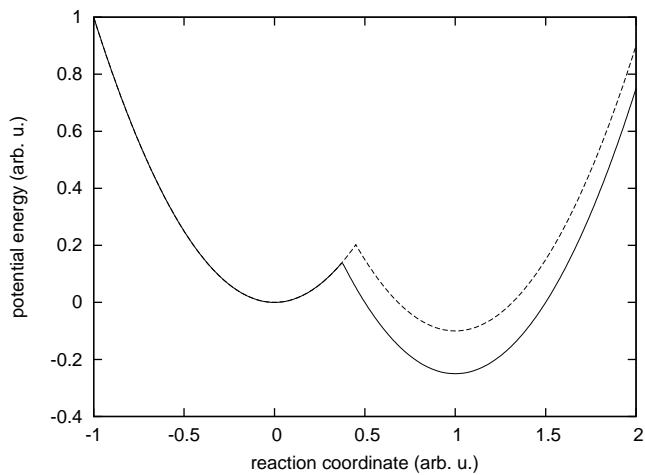

Figure 1: Illustration of the Bell-Evans-Polanyi principle. The potential energy is described by two parabolas. The transition state is the intersection of both parabolas. Evidently the transition state is raised if the local minima on the right hand side is raised (dashed line) and vice versa.

Even though molecular dynamics gives always reasonably low barriers, the parameters $\beta_1$ and *mdmin* can be tuned to vary the barrier height. By reducing the value of $\beta_1$ the system is permitted to perform more unsuccessful escape trials and will thus find on the average lower barriers. By increasing the value of *mdmin* the system can oscillate more frequently within the basin from which it wants to escape or even jump over two barriers and is therefore also more likely to find lower barriers. The last 4 lines of Table 1 show these MD results. The first line of Table 1 was obtained in a run where the escape trials were not done by MD, but by random displacements of the atoms. The step size of the random displacements was also adjusted such that half of all escape trials will lead into new basins. Random displacements can lead to some astronomically large barriers. For this reason $E_{kinetic}$ is in this case not an average barrier, but a typical low barrier height.

All the results show a clear correlation between the average value $< E_{kinetic} >$ and the number of minima that are visited by the simulation before the global minimum is found. This confirms the validity of the Bell-Evans-Polanyi principle. Lower barriers lead on average into lower new basins. Hence higher energy regions that contain a very large number of minima are avoided and the global minimum is found by traversing a smaller number of local minima. Note that our test is particularly stringent because we can not only go down in energy. In order to leave the initial funnel we have to go up in energy as well, but we do not go higher than needed.

Table 1: Correlation between the average barrier height $<E_{kinetic}>$ and the number of minima visited before the global minimum. $<E_{kinetic}>$ is the average of the fluctuating quantity $E_{kinetic}$ over the entire simulation.

| method | $<E_{kinetic}>$ | number of minima visited before global minimum |
|---|---|---|
| random displacement | $\geq 300$ | 34000 |
| MD, $\beta_1 = 1.05$, $mdmin=1$ | 4.7 | 830 |
| MD, $\beta_1 = 1.02$, $mdmin=1$ | 3.6 | 510 |
| MD, $\beta_1 = 1.01$, $mdmin=1$ | 3.2 | 430 |
| MD, $\beta_1 = 1.01$, $mdmin=2$ | .5 | 410 |

The number of minima that are traversed before finding the ground state is subject to large fluctuations. As shown in Fig. 2 this number has a nearly perfect exponential distribution. For this reason the values in Table 1 were averaged over 1000 runs in the case of MD and over 100 runs in the case of random displacements.

Even lower barrier could be obtained by eigenvector following methods [10]. This introduces however some prefered search direction and excludes thus practically some transitions into neighboring basins. In agreement with other studies [11] we therefore found that such choices, that greatly reduce the randomness of the search direction, slow down the search for the global minimum.

# 6 Comparing with similar methods

## 6.1 Basin hopping

Basin hopping has turned out to be a powerful Monte Carlo method for the determination of the global minimum of complex molecular systems. It has been applied to various realistic systems such as silicon clusters [12]. Basin hopping is a method that transforms the potential energy surface in an advantageous way. The value of the potential energy surface within one basin is replaced by the value of the potential energy at the associated minimum as shown in Fig. 3. The common explanation for the success of the basin hopping method is that it eliminates the barriers between the basins of different minima. This is true, but nevertheless, it has to be stressed that the basin hopping method does not eliminate the barriers between super-basins or funnels. The transformed piecewise constant potential energy surface of the basin hopping method still exhibits barriers that have to be overcome by Monte Carlo steps. If the height of these remaining barriers of the transformed surface between super basins is small compared to the height of the original barriers of the untransformed surface between the basins (upper panel of Fig. 3), the basin hopping method is expected to offer a significant advantage, otherwise (lower panel of Fig. 3), the advantage will be marginal. As we have seen, the barrier height encountered in a simulation depends strongly on the kind of moves. Random displacements lead to particularly high barriers.

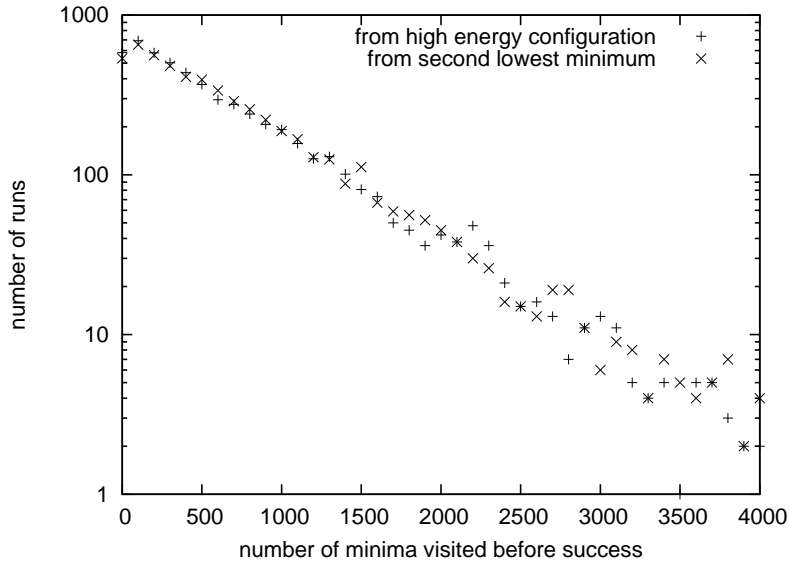

Figure 2: Number of runs that found the global minimum after traversing a certain number of other minima. The results are given both for the case, where the initial configuration is the second deepest minimum and for a high energy starting configuration.

Without introducing a history list of all the previously visited minima, the basin hopping algorithm and the minima hopping algorithm have a rather similar behavior, even though their theoretical principles are different. In contrast to basin hopping, minima hopping is not a Monte Carlo method. What makes the significant difference in practice is the feedback introduced by the history. As a consequence the Minima hopping method can climb out of a 'wrong' funnel much faster than the basin hopping method. It is thus superior to the basin hopping method for systems that have a deep 'wrong' funnel. Wrong means in this context just that the funnel does not contain the global minimum.

Being a thermodynamic method, basin hopping has also the inconvenience that a temperature has to be fixed. Choosing a non-optimal value can prevent finding the global minimum.

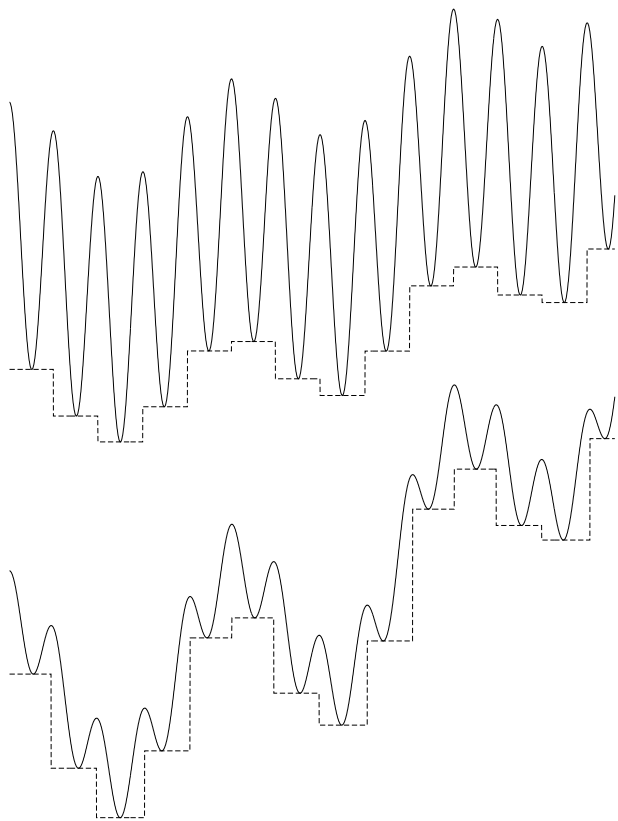

Figure 3: Two different potential energy landscapes: In the upper part the barriers separating basins are very high compared to the energy differences of the various local minima, in the lower part both are comparable.

## 6.2 Activation Relaxation Technique

The Minima hopping method is essentially a sequence of activation relaxation events followed by an acceptance/rejection step. Whereas in the ART method [13] one tries to find very low barriers and in the ART nouveau method [11] even the exact transition state, the Molecular dynamics scheme that we have proposed will in general go over higher barriers. The findings on the validity of the Polyani principle would suggest that ART is even better suited than MD for Minima Hopping. The reason why we have chosen MD instead of ART is that it is not obvious how to introduce a feedback parameter such as $E_{kinetic}$ into the ART scheme. The existence of such a parameter is essential for a history based scheme. Only such a parameter allows us make more violent jumps that explore new regions of the configuration space when the simulation has the tendency to revisit repeatedly certain configurations of the space.

## 6.3 Temperature accelerated molecular dynamics

Temperature accelerated molecular dynamics [14] is a scheme that also uses molecular dynamics to escape from a basin. The temperature is chosen sufficiently high, so that the system will not very long remain in one basin. A modified version of temperature accelerated molecular dynamics [15] also introduces a history that allows for some simplifications if a basin is revisited during the simulation. The primary purpose of temperature accelerated molecular dynamics is however not to find the global minimum, but to follow a system over very long time scales. Therefore temperature accelerated molecular dynamics does not use the history to make more violent moves, that would not have a physical counterpart. Temperature accelerated molecular dynamics also does not have an acceptance/rejection step that gives a preference towards lower energy configurations.

# 7 Numerical tests

Doye and Wales have studied Lennard Jones clusters in great detail and mapped out the structure of their local minima configurations [9]. Their disconnectivity graphs for these systems allow to predict very well whether the global minimum of a system is difficult or easy to find. As a consequence of these detailed investigations the Lennard Jones cluster can be considered as a benchmark system for any new global minimization algorithm. The minima hopping method was therefore also applied to these clusters and the putative global minimum found by Doye and coworkers was easily rediscovered in all cases as shown in Fig. 4 for the 38 atom cluster.

The performance of the minima basin hopping method was also compared with the performance of the basin hopping method using exactly the same type of moves, namely random displacements. For a simple funnel structure like in the 19 atom Lennard Jones cluster the performance of both methods is quite similar. For a system with two deep funnels minima hopping is superior since it climbs out of the wrong funnel much faster. Averaging again over 100 runs for the 38 atom cluster, basin hopping visited some 75000 minima before falling into the global minimum compared to 34000 with minima hopping (Table 1). For the basin hopping run we choose the temperature that gives the fastest success and which corresponds to 1.2 energy units.

To test our algorithms for more difficult systems with more atoms we turned to silicon as described by the EDIP interatomic potential [16]. This system has the advantage that the global minimum is known for any number of atoms. The ground state is the perfect crystal. The simulations were started using an entirely wrong simple cubic crystal structure. With fixed cell size the silicon systems have wrong funnels that contain a very large number of minima. Physically the local minimum at the bottom of a wrong funnel corresponds to a perfect silicon crystal that is rotated with respect to the correct lattice vectors as shown in Fig. 5 for the case of a 64 atom crystal. Rotating the crystal in the correct position is not possible with periodic boundary conditions. So the crystal has to become amorphous before it can it can recrystallize

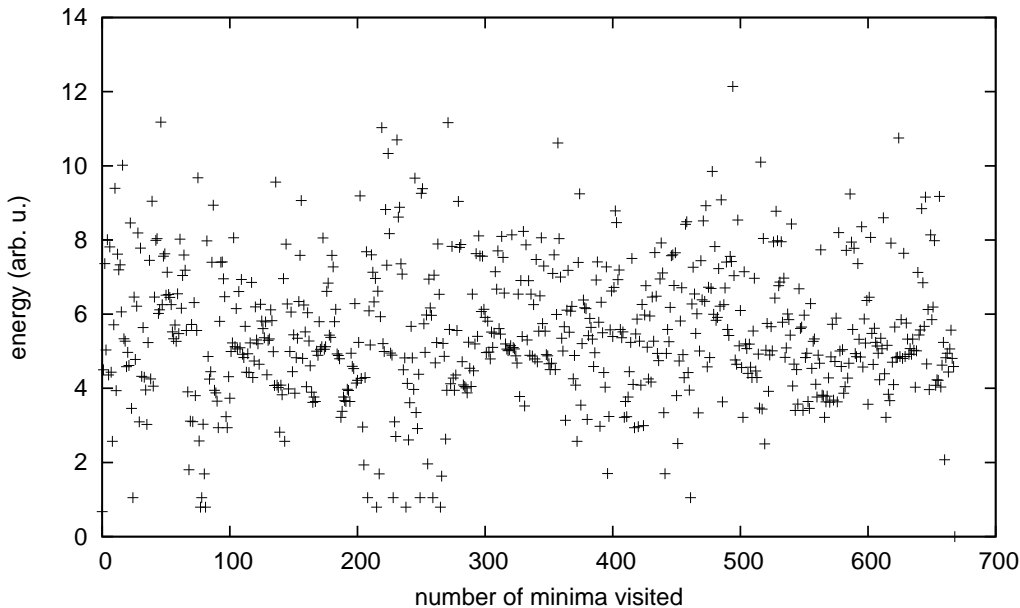

Figure 4: The history of all the minima visited in the search for the ground state of a 38 atom Lennard Jones cluster. The energy was shifted such that the ground state has energy 0.

in the correct position. This requires obviously to overcome very high barriers.

We applied the basin hopping method to 64 and 216 atom crystals and always found the ground state. These system sizes are nearly two orders of magnitude larger than the crystalline systems for which up to now the global minimum was found by simulation [17]. For the silicon systems it also turned out that it is impossible to find the global ground state without our history feedback mechanism. Both basin hopping as well as simulated annealing methods fail for this system. They all get stuck in the astronomically large number of amorphous local minima. Fig 6 illustrates how minima hopping succeeds in finding the global minimum. The algorithm first goes down fairly smoothly. Since it is not difficult to find new lower energy minima both $E_{kinetic}$ and $E_{diff}$ are relatively small. Once the algorithm has arrived at the bottom of a wrong funnel (after some 2.2 million minima), configurations begin to be revisited frequently and $E_{kinetic}$ starts to increase. Because of the violent moves the new configurations are on the average higher in energy and $E_{diff}$ has to increase as well. This is clearly visible as some broadening of the energy history curve. At some point the algorithm has finally climbed out of the wrong funnel and has arrived at the correct funnel. The system then collapses very rapidly into the global ground state after having visited some 4 million minima. Note that this number is very small compared to the total number of minima. Considering only Wooten-Winer-Weaire processes [19], the estimated number of local minima is $2^216 \approx 10^{65}$.

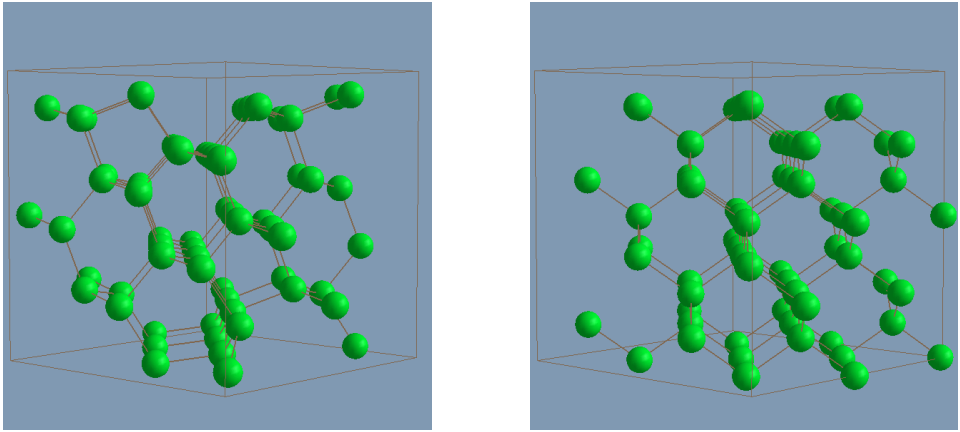

Figure 5: A crystal that is rotated slightly with respect to the ground state of a silicon crystal (left panel) together with the true ground state (right panel).

Many different physical processes can be invoked to cross from one local minimum into another one [20]. The results show that all these processes are activated by the escape trials using molecular dynamics.

## 8  Conclusions

To find the global minimum of a potential energy surface that has multi-funnel structure requires an algorithm that can rapidly climb out of wrong funnels. This feature can only be obtained by abandoning the standard Markov based Monte Carlo methods and by introducing a feedback mechanism, that based on the whole simulation history, enforces the exploration of new regions of the configuration space. The minima hopping contains such a feedback mechanism.

A preference towards low energy configurations is obtained by an acceptance/rejection step. Imposing this preference is simpler if the moves have already a tendency to go into other low energy minima. As expected from the Bell-Evans-Polanyi principle, low energy barriers lead on the average to low energy minima. The molecular dynamics scheme that we use for the escape part gives reasonably low barriers and is applicable to any atomistic system.

## 9  Acknowledgements

I thank Christoph Bruder and Wladimir Hellmann for useful comments on the manuscript and the Swiss National science foundation for their financial support.

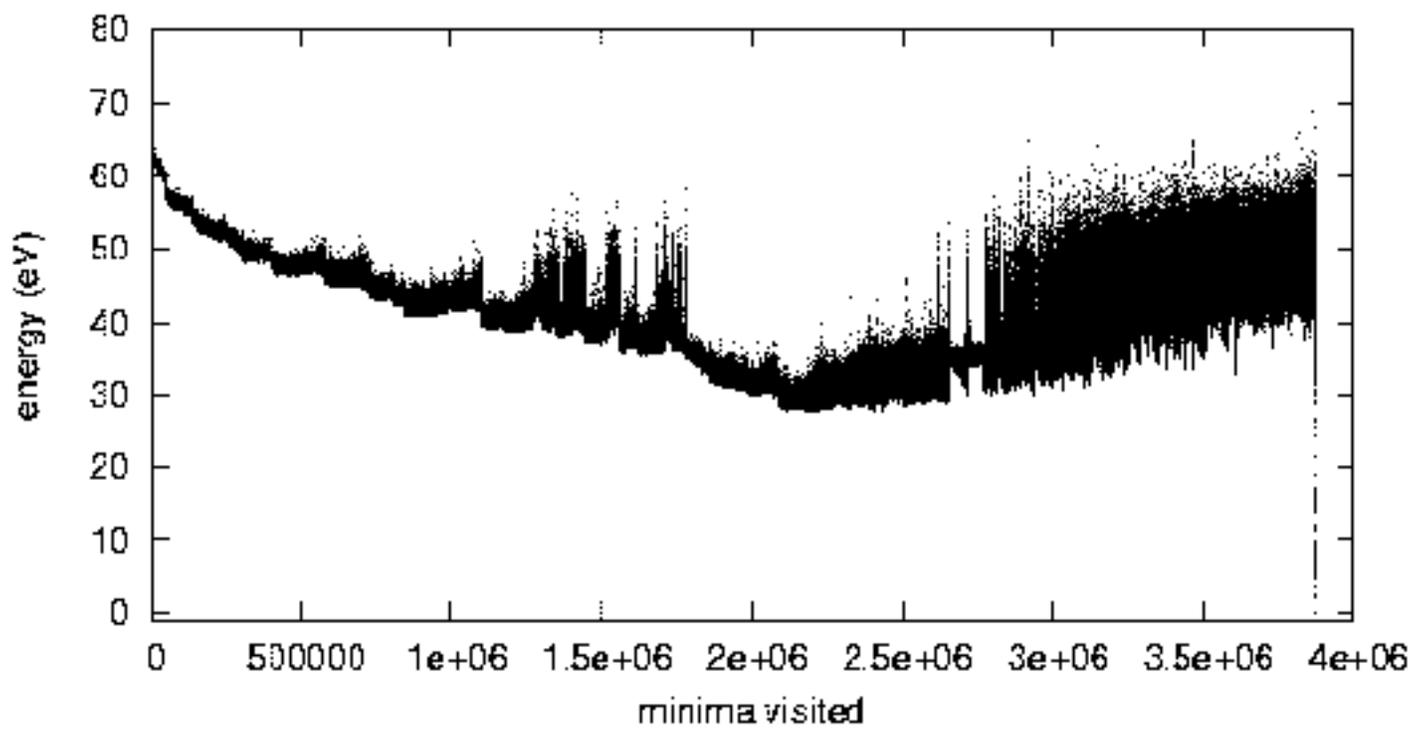

Figure 6: The history of all the minima visited in the search for the ground state of a 216 atom silicon crystal. The energy was shifted such that the ground state has energy 0.

## References

[1] G. Johannesson, T. Bligaard, A. Ruban, H. Skriver, K. Jacobsen, and J. Norskov, Phys. Rev. Lett. 88, 255506 (2002)

[2] S. Chan and K. Dill, Proteins: Structure, Function, and Genetics, **30**, 2 (1998) ; http://laplace.compbio.ucsf.edu/ danny/Protein/fig-gallery.html

[3] D. Goldberg, *Genetic Algorithms in Search, Optimization, and Machine Learning*, Addison Wesley, New York, 1989

[4] P. Salomon, P. Sibani and R. Frost, *Facts, Conjectures and Improvements for Simulated annelaing*, SIAM, Philaelphia (2002).

[5] B. Berg and T. Neuhaus Phys. Lett. B **267**, 249 (1991)

[6]  H. Grubmller Phys. Rev. E **52**, 2893 (1995) ; A. Voter J. Chem. Phys. **106**, 4665 (1997

[7]  A. Laio and M. Parrinello, Proc. Natl. Acad. Sci. U.S.A. **99**, 12 562 (2002)

[8]  F. Jensen, *Computational Chemistry*, Wiley, New York (1999).

[9]  D. Wales M. Miller and J. Doye, J. Chem. Phys. **110** 6896 (1999) ; J. Doye, M. Miller and D. Wales, J. Chem. Phys. **111** 8417 (1999) ;

[10]  L. Munro and D. Wales, Phys. Rev. B **59**, 3969 (1999)

[11]  R. Malek and N. Mousseau, Phys.Rev. E **62**, 7723 (2000)

[12]  S. Yoo and X. Zeng, J. Chem. Phys. **119** 1442 (2003)

[13]  G.T. Barkema and N. Mousseau, Phys.Rev. Lett. **77**, 4358 (1996) ; N. Mousseau and G.T. Barkema, Phys.Rev. E **57**, 2419 (1998)

[14]  M. Sorensen and A. Voter J. Chem. Phys. **112** 9599 (2000)

[15]  F. Montalenti and A. Voter, J. Chem. Phys. **116** 4819 (2002)

[16]  M. Z. Bazant and E. Kaxiras, Phys. Rev. Lett. **77**, 4370 (1996) ; M. Z. Bazant, E. Kaxiras, J. F. Justo, Phys. Rev. B **56**, 8542 (1997) ; J. F. Justo, M. Z. Bazant, E. Kaxiras, V. V. Bulatov, and S. Yip, Phys. Rev. B **58**, 2539 (1998) ; S. Goedecker, Comp. Phys. Commun. 148, **124** (2002)

[17]  D. Williams, Acta Crystallogr A **52** 326 (1996) R. Wawak, J. Pillardy, A. Liwo, K. Gibson and H. Scheraga, J. Phys. Chem. A **102** 2904 (1998) ;

[18]  D. Wales and J. Doye, J. Phys. Chem. A **101** 5111 (1997) ; J. Doye and D. Wales, Phys. Rev. Lett. **80**, 1357 (1998) J. Doye, D. Wales and M. Miller, J. Chem. Phys. **109** 8143 (1998)

[19]  F. Wooten, K. Winer and D. Weaire, Phys. Rev. Lett. **54** 1392 (1985); S. Goedecker, T. Deutsch and L. Billard, Phys. Rev. Lett. **88**, 235501 (2002)

[20]  F. Valiquette and N. Mousseau, Phys. Rev. B **68**, 125209 (2003) ;